\documentstyle[12pt]{article}
\begin{document}

\newcommand{\nc}{\newcommand}

\nc{\rem}[1]{{\bf [#1]}}
\nc{\eqn}[1]{Eq.~(\ref{eq:#1})}
\nc{\eqns}[2]{Eqs.~(\ref{eq:#1}) and (\ref{eq:#2})}
\nc{\fig}[1]{Fig.~\ref{fig:#1}}
\nc{\figs}[2]{Figs.~\ref{fig:#1} and \ref{fig:#2}}
\nc{\be}{\begin{equation}}
\nc{\ee}{\end{equation}}
\nc{\ba}{\begin{array}}
\nc{\ea}{\end{array}}
\nc{\bea}{\begin{eqnarray}}
\nc{\eea}{\end{eqnarray}}
\nc{\beqns}{\begin{eqnarray*}}
\nc{\eeqns}{\end{eqnarray*}}
\nc{\nn}{\nonumber}
%
\input epsf
\nc{\figdir}{} 
\nc{\figmac}[5]{\begin{figure}
\centerline{\parbox[t]{#1in}{\epsfbox{\figdir #2.ps}}}
\caption[#4]{\label{fig:#3} #5}\end{figure}}
%
\def\epsfsize#1#2{\ifdim#1>\hsize\hsize\else#1\fi}

\renewcommand{\slash}[1]{/\kern-7pt #1}

\def\cA{{\cal A}}
\def\cB{{\cal B}}
\def\cC{{\cal C}}
\def\cD{{\cal D}}
\def\cE{{\cal E}}
\def\cF{{\cal F}}
\def\cG{{\cal G}}
\def\cH{{\cal H}}
\def\cI{{\cal I}}
\def\cK{{\cal K}}
\def\cM{{\cal M}}
\def\cN{{\cal N}}
\def\cO{{\cal O}}
\def\cP{{\cal P}}
\def\cR{{\cal R}}
\def\cS{{\cal S}}
\def\cX{{\cal X}}
\def\cZ{{\cal Z}}

\def\qb{\bar{q}}
\def\kb{\bar{k}}
\def\rb{\bar{r}}
\def\Qb{\bar{Q}}
\def\Kb{\bar{K}}
\def\Rb{\bar{R}}

\def\to{\rightarrow}
\nc{\lra}{\leftrightarrow}
\nc{\la}{\langle}
\nc{\ra}{\rangle}
\def\A#1#2{\la#1#2\ra}
\def\B#1#2{[#1#2]}
\def\s#1#2{s_{#1#2}}
\def\h#1#2#3#4{\la#1#2#3#4\ra}
\def\P#1#2{{\cal P}_{#1#2}}
\def\L#1#2{\left\{#2\right\}_{#1}}
\nc{\Lmin}{{\rm L}}
\def\Li{{\rm Li_2}}
\def\ms{$\overline{{\rm MS}}$}
\def\MS{{\overline{{\rm MS}}}}

\nc{\y}{\gamma}
\nc{\yf}{\gamma_5}
\nc{\yh}{\hat{\gamma}}
\nc{\yt}{\tilde{\gamma}}
\nc{\yb}{\bar{\gamma}}
\nc{\ellb}{\bar{\ell}}
\nc{\g}{g}
\nc{\gh}{\hat{g}}
\nc{\gt}{\tilde{g}}
\nc{\qbar}{\bar{q}}
\nc{\ax}{{\rm ax}}
\nc{\Ord}{\cO}

\nc{\al}{\alpha}
\nc{\del}{\delta}
\nc{\eps}{\epsilon}
\nc{\ve}{\varepsilon}
\nc{\ib}{{\bar\imath}}
\nc{\jb}{{\bar\jmath}}

\nc{\kt}{\tilde{k}}
\nc{\Pp}{{\cal P}_+}
\nc{\Pm}{{\cal P}_-}
\nc{\Ppm}{{\cal P}_\pm}
\nc{\Pmp}{{\cal P}_\mp}

\nc{\Nf}{N_f}
\nc{\Nc}{N_c}
\nc{\cg}{c_\Gamma}

\nc{\coeff}[2]{{\textstyle{{#1}\over{#2}}}}
\nc{\hf}{\coeff12}
\nc{\third}{\coeff13}

\nc{\pz}{ {\cal P}_Z(s) }
\nc{\abspz}{ \Bigl\vert \pz \Bigr\vert^2 }
\nc{\repz}{ {\rm Re} \pz }
\nc{\sstw}{ \sin^2\theta_W }
\nc{\stwotw}{ \sin2\theta_W }

\def\pbar{\overline{p}}

\nc{\ycut}{y_{\rm cut}}


\begin{titlepage}
\vspace*{-2cm}
\begin{flushright}
SLAC--PUB--7528\\
hep-ph/9706285\\
June, 1997\\
\end{flushright}
\vskip 2.5cm
\begin{center}
{\Large\bf
Complete $\cO(\alpha_s^3)$ Results for 
$e^+e^- \to (\gamma,Z) \to$ Four Jets
}\footnote{
Research supported by the Department of
Energy under grant DE-AC03-76SF00515, 
and by the Swiss National Science Foundation}\\
\vskip 1cm
{\large  Lance Dixon and Adrian Signer} \\
\vskip 0.3cm
{\it 
Stanford Linear Accelerator Center \\
Stanford University \\
Stanford, CA 94309
} \\
\vskip 2cm
\end{center}

\begin{abstract}
\noindent
We present the next-to-leading order ($\cO(\alpha_s^3)$) 
perturbative QCD predictions for $e^+e^-$ annihilation into four jets.
A previous calculation omitted the $\cO(\alpha_s^3)$ terms
suppressed by one or more powers of $1/\Nc^2$, where $\Nc$ is the number 
of colors, and the `light-by-glue scattering' contributions.
We find that all such terms are uniformly small, 
constituting less than 10\% of the correction.
For the Durham clustering algorithm, the leading and next-to-leading
logarithms in the limit of small jet resolution parameter $\ycut$ 
can be resummed.  We match the resummed results to our fixed-order
calculation in order to improve the small $\ycut$ prediction.
\end{abstract}

\vskip 2cm
\begin{center}
{\sl Submitted to Physical Review D}
\end{center}

\end{titlepage}
\setcounter{footnote}{0}

\bigskip


\section{Introduction}

Electron-positron annihilation into jets provides an arena
for studying quantum chromodynamics (QCD) that is free of
initial-state uncertainties such as parton distribution
functions.  At the large center-of-mass energies achieved by
SLC, LEP, and now LEP2, $e^+e^-$ annihilation is also relatively 
free of nonperturbative final-state effects, {\it i.e.} hadronization
corrections.  On the other hand, perturbative QCD corrections to
jet rates can be very large.  For example, the three-jet rate at 
the $Z^0$ pole receives a 20-30\% correction~\cite{ThreeJetsPrograms}
at order $\alpha_s^2$.  These next-to-leading-order (NLO) corrections 
are of course critical for obtaining a precise experimental measurement 
of $\alpha_s$ from the three-jet rate and related $\cO(\alpha_s)$
observables~\cite{EventShapeAlphas,MatchingExp}.

More recently, the NLO corrections to $e^+e^-$ production of four jets
were computed, and a correction of roughly 100\% was found~\cite{FourJets}
for most jet algorithms (when the renormalization scale was set equal to 
the center-of-mass energy).
This computation omitted terms
suppressed by one or more powers of $1/\Nc^2$, where $\Nc$ is the number 
of colors in a general $SU(\Nc)$ gauge theory ($\Nc=3$ for QCD).
It also neglected the `light-by-glue scattering' contributions ---
interference terms where two different flavor quarks couple to the 
virtual photon or $Z$ boson.  In this article we present the complete 
$\cO(\alpha_s^3)$ results, using an improved version of the same numerical
program, MENLO\_PARC~\cite{MENLOPARC}, 
which was employed for the leading-in-$\Nc$ computation.
The crucial ingredients for the construction of the program
are the tree-level amplitudes for five massless final state partons,
$e^+e^-\to q\qbar ggg$ 
and $e^+e^-\to q\qbar q'\qbar'g$~\cite{BerGie89,FiveJetsBorn},
and especially the recently-computed one-loop virtual amplitudes for 
$e^+e^- \to q\qbar q'\qbar'$~\cite{GloverMiller,Zqqqq}
and $e^+e^- \to q\qbar gg$~\cite{Zqggq}.
We use the formulas given in refs.~\cite{BerGie89,Zqqqq,Zqggq}.

The NLO prediction of the four-jet fraction --- an observable whose 
expansion begins at order $\alpha_s^2$ --- makes it possible 
to measure $\alpha_s^\MS$ with the same formal level of precision (NLO)
as has previously been reserved for $\cO(\alpha_s)$ observables in
$e^+e^-$ annihilation.
However, the theoretical uncertainty in such a measurement will
still be sizeable:
Because the one-loop corrections are so large, the renormalization-scale
dependence of the NLO four-jet result is still strong, and it is 
likely that uncalculated higher-order corrections are important.  
Also, a significant four-jet rate only appears at smaller values
of the jet resolution parameter $\ycut$, where there are large
perturbative logarithms, although these can be partially resummed for 
the Durham algorithm~\cite{DurhamResum}.

There are at least two other motivations for studying $e^+e^-$
annihilation to four jets: (1) These events are a background to 
$e^+e^- \to W^+W^- \to$~4 jets, particularly when the 
center-of-mass energy is not far above the $W$-pair threshold, 
as is the case at LEP2.  (2) Four-jet final states provide 
QCD tests to which three-jet events are insensitive~\cite{FourJetAngles}.  
For example, the non-abelian three-gluon
vertex appears at leading order in four-jet events;
the same is true for the production of hypothetical, light, colored 
but electrically neutral particles, such as light 
gluinos~\cite{eeGluinos,GoMu,OPALAngles,ALEPHAngles}.
In both applications, distributions of the four jets with respect
to energies and angles~\cite{FourJetAngles} are important.
Such distributions can be computed at NLO using the same numerical program, 
and will be the subject of a separate publication~\cite{NLOAngles};
here we briefly study the sensitivity of the total four-jet rate
to additional light fermions.

The remainder of the paper is organized as follows.  In section 2 
we describe the dependence of the four-jet rate on electroweak and
color factors, and outline the structure of the numerical calculation.
In section 3 we present the complete $\Ord(\alpha_s^3)$ predictions
for three different jet algorithms.
We indicate the dependence of the predictions on the (unphysical) 
renormalization scale $\mu$.
The Geneva algorithm~\cite{schemes} has a relatively mild $\mu$ dependence
(small NLO correction) and a relatively strong dependence on the 
number of light quark flavors $\Nf$; we discuss the extent to which 
$\Nf$ can be determined from the Geneva four-jet rate alone.
In section 4 we present results from matching the resummed Durham
jet rate to the fixed-order $\Ord(\alpha_s^3)$ results; the improved
prediction agrees quite well with preliminary SLD data~\cite{SLDdata}.
Section 5 contains our conclusions.


\section{Structure of the Cross-Section and Computation}

For computational reasons as well as to study the effect of varying
parameters, it is useful to decompose the leading-order (Born) 
and NLO contributions to the four-jet differential cross-section 
with respect to both their electroweak and QCD (color) structure.  
To simplify the electroweak decomposition we assume that the observable 
being calculated is insensitive to both 
(1) correlations between the final-state hadrons and the 
electron-positron beam direction, 
and (2) quark and gluon helicities.  
We also assume the positrons are unpolarized and the electrons have 
a longitudinal polarization of $P_e$ ($P_e=+1$ for a right-handed beam).  
QED initial state radiation and other electroweak corrections are 
neglected.  Then the helicity-summed four-jet (differential) 
cross-section at center-of-mass energy $\sqrt{s}$ may be written
\be
\label{eq:ewdecompI}     
\sigma_{4-{\rm jet}} = {4\pi\alpha^2\over3s} \, \Nc \, \left[
  f^{(I)}(s) \, \sigma_4^{(I)} 
+ f^{(II)}(s) \, \sigma_4^{(II)} 
+ f^{(III)}(s) \, \sigma_4^{(III)} \right] \,,
\ee
where
\bea
f^{(I)}(s) &=& \sum_q (Q^q)^2
    + {1\over4} \Bigl((v_L^e)^2(1-P_e)+(v_R^e)^2(1+P_e)\Bigr)
         \sum_q \Bigl( (v_L^q)^2+(v_R^q)^2 \Bigr) \abspz
\nn \\
&& \hskip 1cm
    - {1\over2} \Bigl( v_L^e(1-P_e)+v_R^e(1+P_e) \Bigr)
       \biggl( \sum_q Q^q (v_L^q + v_R^q) \biggr) \repz \, ,
\nn \\
f^{(II)}(s) &=& \biggl(\sum_q Q^q \biggr)^2
    + {1\over8} \Bigl((v_L^e)^2(1-P_e)+(v_R^e)^2(1+P_e)\Bigr)
       \biggl(\sum_q (v_L^q+v_R^q) \biggr)^2 \abspz
\nn \\
&& \hskip 1cm
    - {1\over2} \Bigl( v_L^e(1-P_e)+v_R^e(1+P_e) \Bigr)
       \biggl( \sum_q Q^q \biggr) 
       \biggl( \sum_q (v_L^q + v_R^q) \biggr) \repz \, ,
\nn \\
\label{eq:ewdecompII}     
f^{(III)}(s) &=& 
     {1\over8\sin^2 2\theta_W} \Bigl((v_L^e)^2(1-P_e)+(v_R^e)^2(1+P_e)\Bigr)
       \abspz \, ,
\eea     
where $\alpha$ is the fine structure constant, 
$Q^q$ is the charge of quark $q$ in units of $e$, 
and the left- and right-handed couplings to the $Z^0$ are
\bea
v_L^e &=& { -1 + 2\sin^2 \theta_W \over \sin 2 \theta_W } \,, 
\hskip 2.3 cm 
v_R^e  = { 2 \sin^2 \theta_W \over  \sin 2 \theta_W } \,,  
\nn \\
\label{eq:vlrdef} 
v_L^q &=& { \pm 1 - 2 Q^q\sin^2 \theta_W \over  \sin 2 \theta_W } \,,
\hskip 1.9 cm 
v_R^q = -{ 2 Q^q \sin^2 \theta_W \over \sin 2 \theta_W }  \,, 
\eea
where $\theta_W$ is the weak mixing angle;
the two signs in $v_L^q$ correspond to up $(+)$ and 
down $(-)$ type quarks. 
Equations~(\ref{eq:ewdecompI}) and~(\ref{eq:ewdecompII}) include
both virtual photon and $Z$ boson exchange (and their interference);
the ratio of $Z$ and photon propagators is given by 
\be
\pz = {s \over s - M_Z^2 + i \Gamma_Z \, M_Z} \,,
\ee
where $M_Z$ and $\Gamma_Z$ are the mass and width of the $Z$.  

Representatives of the classes of diagrams contributing to $f^{(I)}$,
$f^{(II)}$ and $f^{(III)}$ 
are depicted in \fig{Contribs} as amplitude interferences.  
Five-parton cuts of these graphs, shown as dashed lines, correspond
to the real part of the NLO correction; four-parton cuts, shown
as dotted lines, correspond to the virtual part.  
In contribution (I) a single fermion couples
to both $(\gamma,Z)$ vector bosons in the interference,
via either a vector or axial vector coupling.  (As shown in the 
figure, there may be a second or even a third fermion loop in the 
interior of the graph, corresponding to `QCD' factors of $\Nf$ in 
the cross-section.)
This contribution dominates the cross-section at $\cO(\alpha_s^2)$ and
as we will see, again at $\cO(\alpha_s^3)$.

The remaining contributions, (II) and (III), have different origins
in the real and virtual parts of the calculation.  
In the real part they come from the $q\qbar q'\qbar' g$ final state when 
the roles of $q$ and $q'$ are exchanged on the opposite side of the cut;
in particular, a different quark pair couples to the $(\gamma,Z)$ on 
each side of the interference.  
In the virtual part they can have the same kind of exchange origin
in $q\qbar q'\qbar'$ final states, but they can also 
arise from either $q\qbar gg$ or $q\qbar q'\qbar'$ graphs where a 
quark loop couples directly to the photon or $Z$ 
(for example the contribution $A_{6;3}$ ($A_6^\ax$) in 
ref.~\cite{Zqqqq}).  

Contribution (II) represents `light-by-glue scattering', 
whereby a different fermion line couples to each vector boson, 
via a vector coupling in each case.  
There is no such contribution at $\cO(\alpha_s^2)$ 
if only charge-blind observables are considered~\cite{ERT}, due to 
Furry's theorem --- the order $\alpha_s^2$ amplitude interferences
all contain fermion triangle subgraphs.  
Although the cross-section is nonvanishing at
$\cO(\alpha_s^3)$, we shall see that it is still extremely small, 
due partly to cancellations in the sum over quark couplings in $f^{(II)}(s)$,
and partly to approximate cancellations in the phase-space integrations
that are related to the exact cancellations at order $\alpha_s^2$. 

Contribution (III), `$Z$-by-glue' scattering, 
is similar to contribution (II) except that the
quarks couple to the $Z$ through the axial vector coupling.  This 
contribution is nonzero at $\cO(\alpha_s^2)$~\cite{AxialCoupling}, 
although small for the three- and four-jet rates, and it remains small at
$\cO(\alpha_s^3)$.  In \eqn{ewdecompII} we have already carried out the 
sum over the five light quark flavors, in which the massless weak isospin 
doublets $(u,d)$ and $(c,s)$ cancelled, leaving only the
$(t,b)$ contribution.  
The top quark contribution to (III) is purely virtual for 
$\sqrt{s} < 2 m_t$, but it does not decouple in the large $m_t$ 
limit~\cite{AxialCoupling}.  We expand in the limit of large top quark
mass, including all terms through $\cO(s/m_t^2)$; at this order
the top quark does not appear in the vector contribution 
(II)~\cite{Zqqqq,Zqggq}.

 
\figmac{6.5}{contribs}{Contribs}{} 
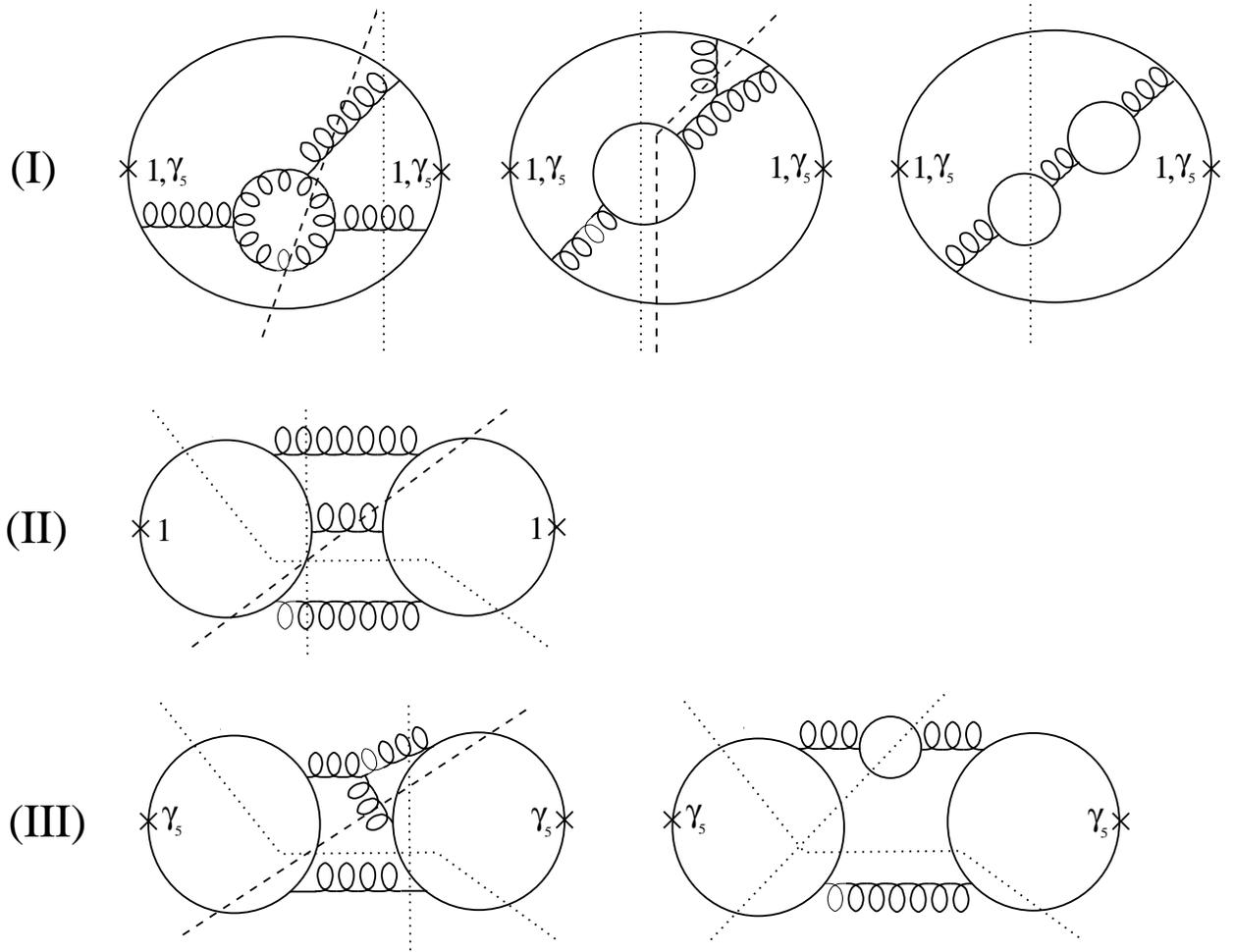
{{\small Representative contributions of type (I), (II) and (III),
as described in the text.  
The coupling of a quark to the $(\gamma,Z)$ vector boson 
is denoted by $\times$, with a $1$ ($\gamma_5$) for vector 
(axial vector) coupling.  Dashed lines correspond to representative
five-parton cuts; dotted lines to four-parton cuts.
\hfill}}


Dividing the four-jet cross-section $\sigma_{4-{\rm jet}}$ by
the total hadronic cross-section at $\cO(\alpha_s)$,
\be
\sigma_{{\rm tot}} = {4\pi\alpha^2\over3s} \Nc \, f^{(I)}(s) 
   \left( 1 + {\alpha_s\over\pi} \right) \,,
\ee
yields the four-jet fraction
\be
R_4 \equiv { \sigma_{4-{\rm jet}} \over \sigma_{{\rm tot}} } 
= \left[ \sigma_4^{(I)} + \frac{f^{(II)}}{f^{(I)}} \sigma_4^{(II)}
+ \frac{f^{(III)}}{f^{(I)}} \sigma_4^{(III)}  \right] \,
 \left( 1 + {\alpha_s\over\pi} \right)^{-1} \,.
\ee
Neglecting for the moment the renormalization-scale dependence of the
calculated cross-section we write the expansion in $\alpha_s$ as
\bea
R_4 &=& \Biggl[ \left( \frac{\alpha_s}{2\pi} \right)^2 B_4
+ \left( \frac{\alpha_s}{2\pi} \right)^3 C_4 \Biggr] 
 \, \left( 1 + {\alpha_s\over\pi} \right)^{-1}  \,,
\label{eq:R4expansion} \\
{ f^{(A)}\over f^{(I)} } \, \sigma_4^{(A)} &=& 
  \left( \frac{\alpha_s}{2\pi} \right)^2 B_4^{(A)}
+ \left( \frac{\alpha_s}{2\pi} \right)^3 C_4^{(A)} \,,
\qquad A=I,II,III,
\label{eq:sig4expansion}
\eea
Next we decompose the one-loop correction to $\sigma_4^{(I)}$ with 
respect to $\Nc$ and $\Nf$:
\be  \label{eq:decompI}
 C_4^{(I)} =
\Nc^2 (\Nc^2 - 1) \Biggl[
 \sigma_4^{(a)} + {\Nf\over\Nc} \sigma_4^{(b)} 
 + {\Nf^2\over\Nc^2} \, \sigma_4^{(c)} 
 + {1\over\Nc^2} \sigma_4^{(d)}  + {\Nf\over\Nc^3} \sigma_4^{(e)} 
 + {1\over\Nc^4} \sigma_4^{(f)}  \Biggr] \,.
\ee
Correspondingly, we write the full $\cO(\alpha_s^3)$ correction to 
the four-jet rate as
\be  \label{eq:decompC}
C_4 = C_4^{(a)} + C_4^{(b)} + C_4^{(c)} + 
          C_4^{(d)} + C_4^{(e)} + C_4^{(f)}
      + C_4^{(II)} + C_4^{(III)} \,,
\ee
absorbing all prefactors into the definitions of the $C_4^{(x)}$.
In ref.~\cite{FourJets} we calculated $C_4^{(a,b,c)}$;
here we add the remaining terms in \eqn{decompC}.
The subleading-color terms $C_4^{(d,e,f)}$ come partly
from non-planar interference graphs (not shown in \fig{Contribs}).
They include identical-quark Pauli exchange contributions
analogous to the E terms of ref.~\cite{ERT}, as well as 
various subleading-color virtual subamplitudes~\cite{Zqqqq,Zqggq},
and subleading terms in the real and virtual color sums.
We find that all the additional terms are considerably smaller
than $C_4^{(a,b,c)}$, at least for the overall four-jet rate.

The Monte Carlo integrations required to numerically evaluate 
the $C_4^{(x)}$ are done separately for each term, except that
$C_4^{(d)}$ and $C_4^{(f)}$ are combined.
An advantage~\cite{FourJets} of breaking up the problem in this way is that  
the $1/\Nc^2$-suppressed integrands have significantly more complicated
analytic representations than the leading terms, and therefore
take more time per point to evaluate (in some cases up to a factor of
five longer).  On the other hand, the $1/\Nc^2$
parametric suppression implies that far fewer numerical evaluations 
of the subleading terms are required in order to achieve an absolute 
statistical accuracy comparable to that for the leading-in-$\Nc$ terms.  
Contributions (II) and (III) could have been further decomposed by
analogy to \eqn{decompI}, but in view of their small overall 
contribution they were each integrated as a single expression. 

As in any NLO QCD computation, the real and virtual corrections to the 
cross-section are separately divergent, but have a finite sum.
In dimensional regularization with $D=4-2\epsilon$, the singularities 
of the virtual part manifest themselves as poles in $\epsilon$
in the one-loop amplitudes, whereas the real singularities are 
obtained upon phase-space integration of the squared tree amplitudes.
We use a general version of the subtraction method~\cite{ERT} to
extract the singular parts of the real cross-section and combine
them with the virtual poles.  This method leaves a finite integral over 
five-parton phase space, and another over four-parton phase space,  
which are performed by adaptive Monte Carlo integration using
VEGAS~\cite{VEGAS}.
The particular form of the subtraction method used here is
essentially that described in ref.~\cite{FKS}, 
to which we refer the reader for more details. 
No approximation of the matrix elements or the phase-space 
has to be made in this method.

The subtraction method relies on the fact that the integral over the
tree cross-section is rendered finite by subtracting all soft and collinear
limits. This means that for a phase-space point that lies very close to a
singular point, the integrand is the square of the difference of two 
large numbers, namely the tree amplitude and its soft or collinear limit. 
In order to obtain the desired cancellation it is crucial to compute 
this difference in a numerically stable way, even if a certain invariant 
mass becomes very small.   Thus, if the phase-space point is so close to 
a singular point that a straightforward evaluation of the amplitude
becomes unstable, the amplitude is replaced by its (more stable)
soft or collinear limit.
We checked that the error introduced by this treatment is 
completely negligible. 
We also checked that our results are independent of the arbitrary
parameters $\delta$ and $\xi_{cut}$ which have to be introduced in
the subtraction method~\cite{FKS}. 

Another potential numerical problem is related to spurious singularities
in the one-loop amplitudes.  Besides the expected poles in the soft and
collinear limits (which are avoided by the program since they lie in the
three-jet region), the one-loop amplitudes have unphysical poles, 
{\it i.e.} poles with zero residue.  Unfortunately, it is not possible to
eliminate all these poles analytically, as long as the amplitude is
expressed in terms of logarithms and dilogarithms multiplied by
kinematic coefficients~\cite{Zqggq}; this elimination is only possible 
if the amplitude is rewritten in terms of more general 
functions~\cite{CamGlMi96}.  
However, in the helicity formalism, one can simplify the (di)logarithmic
coefficients to greatly alleviate the spurious poles~\cite{Zqggq}.
We checked that the numerical evaluation of the matrix elements as given in
refs.~\cite{Zqqqq,Zqggq} is stable, even for points that are quite close to
a spurious pole, and that the probability for hitting an unstable point in
the Monte Carlo integration is very small.  Indeed, we had to evaluate
close to a million points in a test run (corresponding to sub-percent
statistical accuracy on the integral) in order to find one point that was
`close' to a particular spurious pole; at that one point the value of the
vanishing denominator was still about an order of magnitude larger than
where the numerical evaluation of the cross-section typically becomes
unstable.


\section{Fixed-order Results}

We now present results for the four-jet fraction $R_4$
at next-to-leading order in $\alpha_s$.  We use
$\Nc=3$ colors, $\Nf=5$ massless quarks, a strong coupling
constant of $\alpha_s(M_Z) = 0.118$, 
a top mass of $m_t = 175$~GeV,
a $Z^0$ mass and width of $M_Z = 91.187$~GeV and $\Gamma_Z = 2.490$~GeV,
and a weak mixing angle of $\sstw = 0.230$~\cite{ParticleDataGroup}.
The numerical results given here are for $\sqrt{s}=M_Z$, 
but to the extent that contributions (II) and (III) can be neglected,
$R_4$ depends essentially only on $\Nc$, $\Nf$ and $\alpha_s(\sqrt{s})$.
We consider the E0, Durham~\cite{Durham,DurhamResum} 
and Geneva~\cite{schemes} jet algorithms.
These cluster algorithms begin with a set of final-state particles 
(partons in the QCD calculation) and cluster the pair $\{i,j\}$ with the 
smallest value of a dimensionless measure $y_{ij}$ into a single
``proto-jet''.  The procedure is repeated until all the $y_{ij}$ 
exceed the value of the jet resolution parameter $\ycut$, at which
point the proto-jets are declared to be jets.
The algorithms differ in the measure $y_{ij}$ used and/or in the 
rule used to assign a four-momentum $p_{ij}$ to 
two clustered momenta $p_i$, $p_j$.
The same value of $\ycut$ in different schemes may sample 
quite different classes of events. 
For the reader's convenience, we collect the definitions of the 
E0, Durham and Geneva schemes in Table~\ref{tab:jetalgs}.


\begin{table}
\caption{Jet algorithm definitions} 
\label{tab:jetalgs}
\begin{center}
\begin{tabular}{|c|c|c|l|}
\hline
Algorithm&$y_{ij}$&$p_{ij}$\\
\hline
E0& $\frac{(p_i+p_j)^2}{s}$ &
$(E_i+E_j) (1, \frac{{\vec p}_i+{\vec p}_j}{|{\vec p}_i+{\vec p}_j|} )$\\
Durham&2 min$(E_i^2,E_j^2) \frac{1-\cos\theta_{ij}}{s} $& $p_i+p_j$ \\
Geneva&$\coeff89 E_i E_j \frac{1-\cos\theta_{ij}}{(E_i+E_j)^2}$ & 
                 $p_i+p_j $\\
\hline
\end{tabular}
\end{center}
\end{table}

We start the presentation of the results with the E0 scheme. \fig{e0}a
shows the absolute value of the contributions of the different 
electroweak/color pieces to the four-jet fraction at  $\sqrt{s} = M_Z$,
as a function of $\ycut$, setting the renormalization scale to 
$\mu = M_Z$.   Note (from Table~\ref{tab:E0results}) that  
$C_4^{(b)}+C_4^{(c)}$, $C_4^{(d)}+C_4^{(f)}$ and $C_4^{(III)}$ are
negative.   
These curves are compared to preliminary SLD data points~\cite{SLDdata} 
which have been corrected for detector effects and hadronization.  
Obviously the comparison would benefit from a re-analysis using the 
full current $Z^0$ pole data samples. 
As expected, the subleading-color pieces are roughly 10\% 
of the corresponding leading-color contributions,
reflecting the $1/\Nc^2$ suppression.  This feature holds separately for
the terms lacking and having an $\Nf$ factor.  
The contributions (II) and (III) are
so small that we multiply them by a factor of 1000 and 10 respectively 
in the figure. 
Table~\ref{tab:E0results} presents the same results, 
for $\ycut \in \{ 0.005, 0.01, 0.03 \}$,
namely the coefficients 
$(\alpha_s/2 \pi)^3 \, C_4^{(x)}/(1 + \frac{\alpha_s}{\pi})$ 
at $\sqrt{s}=M_Z$, including the statistical uncertainties 
from Monte Carlo integration.  The `Born' line gives the tree-level
result $(\alpha_s/2 \pi)^2 \, B_4/(1 + \frac{\alpha_s}{\pi})$.

 
\figmac{5}{E0}{e0}{} 
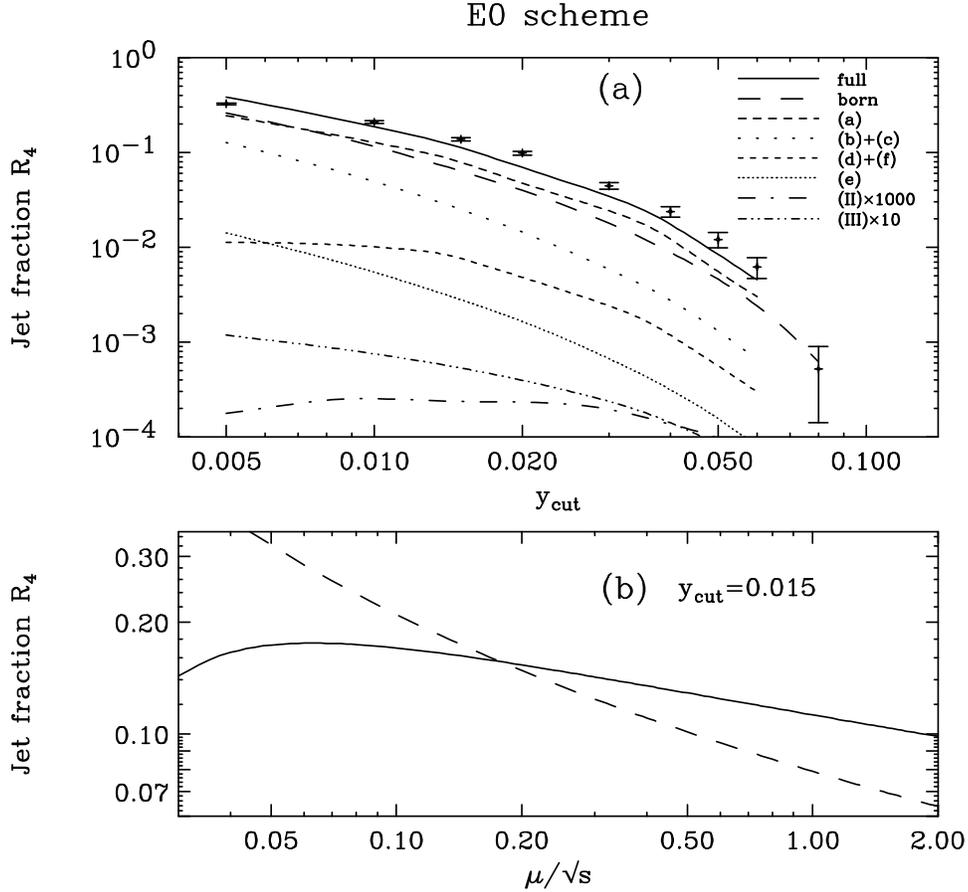
{{\small (a) Absolute value of the contributions of the different
electroweak/color pieces to the four-jet fraction at  $\sqrt{s} = M_Z$ for
the E0 scheme, {\it i.e.} 
$(\alpha_s/2 \pi)^3 \, |C_4^{(x)}|/(1 + \frac{\alpha_s}{\pi})$ with 
$ x \in \{ a,b,c,d,e,f,II,III \}$.  We also show the Born and full
one-loop prediction, and data from ref.~\cite{SLDdata}.  
(b) Dependence of the tree-level (dashed line) and one-loop (solid
line) prediction on the renormalization scale $\mu$ for $\ycut = 0.015$. 
\hfill}}



\begin{table}
\caption{E0 algorithm} 
\label{tab:E0results}
\begin{center}
\begin{tabular}{|c|r|r|r|}
\hline
Contribution to $R_4$ &$\ycut=0.005 \qquad $&$\ycut=0.01\qquad $
&$\ycut=0.03 \qquad $\\
\hline
Born &$ (2.60\pm 0.02) \cdot 10^{-1} $&$(1.16\pm 0.01 ) \cdot 10^{-1} $&
$(1.79\pm 0.01 ) \cdot 10^{-2}$\\
\hline
a  &$(2.43\pm 0.08) \cdot 10^{-1} $& $(1.27\pm 0.03) \cdot 10^{-1}$&  
$(2.42\pm 0.05) \cdot 10^{-2} $\\
b  & $- (1.23\pm 0.02) \cdot 10^{-1} $&$-(4.75 \pm 0.04) \cdot 10^{-2}$&  
$-(5.57 \pm 0.06) \cdot 10^{-3} $\\
c  &$-(4.06 \pm 0.02) \cdot 10^{-3}$&$-(1.83\pm 0.01) \cdot 10^{-3}$& 
 $-(2.93\pm0.01) \cdot 10^{-4}$\\
d+f  &$-(1.13\pm 0.18) \cdot 10^{-2}$&$-(1.01\pm 0.08) \cdot 10^{-2}$& 
 $-(2.42\pm 0.10) \cdot 10^{-3}$\\
e  &$(1.42\pm 0.01) \cdot 10^{-2}$&$(5.45\pm 0.04) \cdot 10^{-3}$&  
 $(6.69\pm 0.06) \cdot 10^{-4}$\\
II  &$(1.66\pm 0.28) \cdot 10^{-7} $&$(2.43\pm 0.32) \cdot 10^{-7} $& 
 $(1.88 \pm 0.18) \cdot 10^{-7} $\\
III  &$- (1.18\pm 0.01) \cdot 10^{-4} $&$-(7.53\pm 0.03) \cdot 10^{-5} $ &
$-(2.37\pm 0.02) \cdot 10^{-5} $\\
\hline
Full$\,\equiv R_4$
 &  $(3.79 \pm 0.08) \cdot 10^{-1} $&$(1.88\pm 0.03 ) \cdot 10^{-1} $&
$ (3.46\pm 0.05 ) \cdot 10^{-2} $ \\
\hline
\end{tabular}
\end{center}
\end{table}

Observable quantities calculated in QCD should be independent of the
arbitrary renormalization scale $\mu$.  However, the perturbative 
expansion is invariably truncated at a finite order, leading
to a residual dependence of the result on $\mu$.
The tree-level $\mu$ dependence is much stronger for the four-jet rate 
than for the three-jet rate, because the former is proportional to
$\al_s^2$ instead of $\al_s$.  The full $\mu$-dependence of the 
NLO four-jet rate is given by
\be
\label{eq:alphadecompmu}
\sigma_4(\mu) = 
 \biggl({\alpha_s(\mu)\over2\pi}\biggr)^2 B_4
+ \biggl({\alpha_s(\mu)\over2\pi}\biggr)^3 \biggl[ 
   C_4 + 2 \beta_0 \ln\biggl({\mu^2\over s}\biggr) B_4 \biggr] \, ,
\ee
where $\alpha_s(\mu)$ is the two-loop running coupling, 
\bea
\alpha_s(\mu) &=& {\alpha_s(M_Z) \over w}
     \left( 1 - {\alpha_s(M_Z) \over \pi} {\beta_1\over\beta_0} 
                {\ln(w)\over w} \right) \, , 
\nn \\
\label{eq:twoloopalpha}
w &=& 1 - \beta_0 {\alpha_s(M_Z) \over \pi} 
        \ln\biggl({M_Z\over\mu}\biggr) \, ,
\eea
with  
$\beta_0 = \coeff{1}{2} (\coeff{11}{3}C_A - \coeff{2}{3} N_f)$,
$\beta_1 = \coeff{1}{4} (\coeff{17}{3}C_A^2-(\coeff{5}{3}C_A+C_F)N_f)$, 
$C_A = \Nc$, $C_F = (\Nc^2-1)/(2\Nc)$. 
As expected, the strong renormalization-scale dependence of the 
tree-level result is reduced
by the inclusion of the next-to-leading order contribution.   
\fig{e0}b plots the $\mu$-dependence of $R_4$ at tree-level
and at one-loop for the E0 scheme, at $\ycut = 0.015$. 

The results for the Durham scheme are presented in 
Table~\ref{tab:Durhamresults}, for the same values of $\ycut$ 
as in the E0 scheme.  Again, the subleading-color terms are 
of the expected size.


\begin{table}
\caption{Durham algorithm} 
\label{tab:Durhamresults}
\begin{center}
\begin{tabular}{|c|r|r|r|}
\hline
Contribution to $R_4$ &$\ycut=0.005\qquad $&$\ycut=0.01 \qquad $
&$\ycut=0.03 \qquad $\\
\hline
Born &$ (6.78\pm 0.02) \cdot 10^{-2} $&$(2.87\pm 0.01 ) \cdot 10^{-2} $&
$(4.11\pm 0.01 ) \cdot 10^{-3}$\\
\hline
a  &$(6.60\pm 0.13) \cdot 10^{-2} $& $(3.03\pm 0.06) \cdot 10^{-2}$&  
$(4.23\pm 0.07) \cdot 10^{-3} $\\
b  & $- (2.68\pm 0.02) \cdot 10^{-2} $&$-(1.03 \pm 0.01) \cdot 10^{-2}$&  
$-(1.24 \pm 0.02) \cdot 10^{-3} $\\
c  &$-(1.27 \pm 0.01) \cdot 10^{-3}$&$-(5.16\pm 0.02) \cdot 10^{-4}$& 
 $-(6.94\pm0.02) \cdot 10^{-5}$\\
d+f  &$-(4.54\pm 0.41) \cdot 10^{-3}$&$-(2.50\pm 0.09) \cdot 10^{-3}$& 
 $-(3.67\pm 0.45) \cdot 10^{-4}$\\
e  &$(2.93\pm 0.02) \cdot 10^{-3}$&$(1.14\pm 0.01) \cdot 10^{-3}$&  
 $(1.43\pm 0.01) \cdot 10^{-4}$\\
II  &$(2.28\pm 0.20) \cdot 10^{-7} $&$(2.22\pm 0.12) \cdot 10^{-7} $& 
 $(9.06 \pm 0.39) \cdot 10^{-8} $\\
III  &$- (5.57\pm 0.03) \cdot 10^{-5} $&$-(3.16\pm 0.02) \cdot 10^{-5} $ &
$-(7.82\pm 0.07) \cdot 10^{-6} $\\
\hline
Full$\,\equiv R_4$
 &  $(1.04 \pm 0.02) \cdot 10^{-1} $&$(4.70\pm 0.06 ) \cdot 10^{-2} $&
$ (6.82\pm 0.08 ) \cdot 10^{-3} $ \\
\hline
\end{tabular}
\end{center}
\end{table}


\begin{table}
\caption{Geneva algorithm} 
\label{tab:Genevaresults}
\begin{center}
\begin{tabular}{|c|r|r|r|}
\hline
Contribution to $R_4$ &$\ycut=0.02\qquad $&$\ycut=0.03 \qquad $
&$\ycut=0.05 \qquad $\\
\hline
Born &$ (2.63\pm 0.02) \cdot 10^{-1} $&$(1.50\pm 0.01) \cdot 10^{-1} $&
$(6.33\pm 0.02) \cdot 10^{-2}$\\
\hline
a  &$(1.16\pm 0.05) \cdot 10^{-1} $& $(8.91\pm 0.25) \cdot 10^{-2}$&  
$(4.90\pm 0.14) \cdot 10^{-2} $\\
b  & $- (1.37\pm 0.02) \cdot 10^{-1} $&$-(6.99\pm 0.09) \cdot 10^{-2}$&  
$-(2.51\pm 0.03) \cdot 10^{-2} $\\
c  &$-(7.78\pm 0.12) \cdot 10^{-3}$&$-(4.32\pm 0.04) \cdot 10^{-3}$& 
 $-(1.68\pm 0.02) \cdot 10^{-3}$\\
d+f  &$(6.90\pm 1.07) \cdot 10^{-3}$&$-(1.10 \pm 1.88) \cdot 10^{-3}$& 
 $-(2.55 \pm 0.78 ) \cdot 10^{-3}$\\
e  &$(1.44\pm 0.02) \cdot 10^{-2}$&$(7.58\pm 0.08) \cdot 10^{-3}$&  
 $(2.83\pm 0.03) \cdot 10^{-3}$\\
II  &$(1.72\pm 0.52) \cdot 10^{-7} $&$(2.89\pm 0.47) \cdot 10^{-7} $& 
 $(2.53\pm 0.35 ) \cdot 10^{-7} $\\
III  &$- (1.06\pm 0.02) \cdot 10^{-4} $&$-(7.86\pm 0.06 ) \cdot 10^{-5} $ &
$-(4.91\pm 0.04) \cdot 10^{-5} $\\
\hline
Full$\,\equiv R_4$
 &  $(2.56\pm 0.06) \cdot 10^{-1} $&$(1.71\pm 0.03) \cdot 10^{-1} $&
$ (8.58\pm 0.15) \cdot 10^{-2} $ \\
\hline
\end{tabular}
\end{center}
\end{table}

The Geneva algorithm has the feature that the leading-order
results, evaluated at $\mu=\sqrt{s}$, give a reasonable description 
of the data for large values of $\ycut$, 
although the shape of the prediction is not quite correct,
especially at small $\ycut$.  Also, the renormalization-scale dependence
is quite flat at moderate $\ycut$.   Finally, the dependence of the
prediction on the number of light flavors $\Nf$ is reasonably large, 
at least in comparison with other algorithms (see
Table~\ref{tab:Genevaresults}).  There is some interest 
in experimentally constraining $\Nf$, in particular because a massless 
gluino would effectively shift the value of $\Nf$ by 
$\Delta\Nf=+3$ in $\cO(\alpha_s^2)$ 
four-jet distributions~\cite{eeGluinos,GoMu}. 
(At $\cO(\alpha_s^3)$ the effect is not simply given by $\Delta\Nf=+3$,
as is illustrated by the structure of the $\cO(\alpha_s^3)$ results for 
the total $e^+e^-$ hadronic cross-section~\cite{eeGluinoTotal}.)
Various authors have suggested that the existence of a light gluino
is already in doubt~\cite{GoMu,CsFo,ALEPHAngles}. 
Nevertheless, we would like to ask whether one
can determine $\Nf$ with sufficient accuracy solely from the overall 
four-jet rate in the Geneva algorithm.  
In \fig{genf} we plot the NLO Geneva prediction as a function
of $\ycut$ for $\Nf=5$ ($u,d,s,c$ and $b$ quarks) and $\Nf=8$
($u,d,s,c$ and $b$ quarks, plus a massless gluino), 
where the bands represent the variation of $\mu$ over the interval 
$[\hf\sqrt{s},2\sqrt{s}]$ and $[\third\sqrt{s},3\sqrt{s}]$ respectively.  
These bands are compared to preliminary SLD data~\cite{SLDdata}. 
The huge uncertainty for small values of
$\ycut$ reflects the fact that the fixed-order prediction is not
converging well for $\ycut \le 0.02$, presumably due to large logarithms 
of $1/\ycut$.
This breakdown happens at larger $\ycut$ for $\Nf = 8$, since in
this particular case $C_4^{(b)}$ is the dominant contribution to the
one-loop correction, and it is further enhanced if $\Nf$ is increased 
from 5 to 8.

As can be seen in \fig{genf} the data tend to favor $\Nf = 5$, at least
for $0.03 \le \ycut \le 0.04$, however, the  uncertainties coming from
uncalculated higher-order terms are still too large to permit excluding
light gluinos using this observable alone.

 
\figmac{5}{GeNf}{genf}{} 
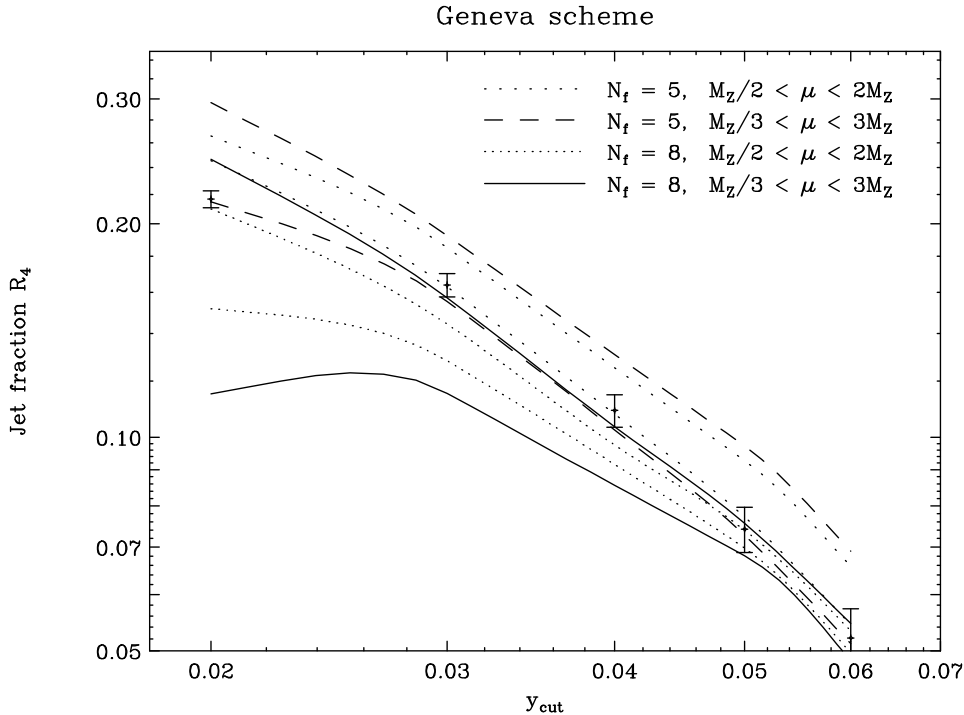
{{\small NLO prediction for the four-jet rate using the Geneva algorithm 
for $\Nf = 5$ and $\Nf = 8$.  The theoretical bands have been obtained 
by varying the renormalization scale from $\hf \sqrt{s} < \mu < 2\sqrt{s}$
and from $\third\sqrt{s} < \mu < 3\sqrt{s}$. 
The data are from ref.~\cite{SLDdata}.
\hfill}}


Various angular distributions in four-jet events
have been proposed to help separate the relatively small contributions
of four-quark final states from the dominant two-quark two-gluon
final states~\cite{FourJetAngles}.  These distributions have been
studied at leading order in $\alpha_s$ in order to constrain $\Nf$ 
as well as the other color factors 
$C_A$ and $C_F$~\cite{AngleExp,OPALAngles,ALEPHAngles}.  
The next-to-leading-order corrections to the distributions 
will be discussed elsewhere~\cite{NLOAngles}, but they are remarkably
small, given the size of the corrections to the overall four-jet rate.
Unfortunately, in many cases the dependence on $\Nf$ is not that
strong, such that a precise determination of $\Nf$ is difficult
in the face of hadronization uncertainties.


\section{Resummed Results}

The four-jet fraction declines rapidly at large $\ycut$,
and there is little data publicly available with 
which to compare our predictions for $\ycut > 0.07$.  
On the other hand, at the kinematic limit $\ycut \to 0$ 
the QCD expansion parameter becomes $\alpha_s L^2$, where 
$L = \ln(1/\ycut)$, and the NLO prediction would be improved if
these large logarithms could be resummed.  
This is possible at leading order (LL) and next-to-leading order (NLL) 
in $L$ in the Durham clustering algorithm because the phase space 
factorizes appropriately~\cite{DurhamResum}.  
The NLL four-jet rate is then given by~\cite{DurhamResum}
\bea
\label{eq:NLLrate}
\hskip -0.3cm   
R_4^{\rm NLL} &=& 2 \, \bigl[ \Delta_q(Q) \bigr]^2 
  \biggl[ \Bigl( \int_{Q_0}^Q dq \, \Gamma_q(Q,q) \Delta_g(q) \Bigr)^2
\nn \\
&& \hskip 2cm
    +  \int_{Q_0}^Q dq \, \Gamma_q(Q,q) \Delta_g(q)
          \int_{Q_0}^q dq' \, 
    \bigl( \Gamma_g(q,q') \Delta_g(q') + \Gamma_f(q') \Delta_f(q') \bigr)
  \biggr] \,.  
\eea
The NLL emission probabilities are
\bea
  \Gamma_q(Q,q) &=& {2C_F\over\pi} {\alpha_s(q)\over q} 
        \left( \ln{Q\over q} - {3\over4} \right) \, ,
\nn \\
  \Gamma_g(Q,q) &=& {2C_A\over\pi} {\alpha_s(q)\over q} 
        \left( \ln{Q\over q} - {11\over12} \right) \, , \nn \\
\label{eq:NLLemit}   
  \Gamma_f(q) &=& {\Nf\over 3 \pi} {\alpha_s(q)\over q}  \, ,
\eea
and the Sudakov factors (probability of no emission) are
\bea
  \Delta_q(Q) &=& 
    \exp\left( - \int_{Q_0}^Q dq \, \Gamma_q(Q,q) \right) \, ,
\nn \\
  \Delta_g(Q) &=& 
    \exp\left( - \int_{Q_0}^Q dq \, 
        \bigl[ \Gamma_q(Q,q) + \Gamma_f(q) \bigr] \right) \, ,
\nn \\
\label{eq:Sudakov} 
   \Delta_f(Q) &=& { \bigl[ \Delta_q(Q) \bigr]^2 
                        \over \Delta_g(Q) } \, .    
\eea

The Durham four-jet rate is an example of a quantity that can
be resummed at leading and next-to-leading logarithmic order, 
but which does not exponentiate.  The NLL results for such 
quantities do not include the proper renormalization-scale dependence
of even the leading-log terms~\cite{QCDColliderBook}:  
Under a change of renormalization scale,
a leading term $\alpha_s^n L^{2n}$ varies by 
$\sim \alpha_s^{n+1} L^{2n} = \alpha_s^{n+1} L^{2(n+1)-2}$,
which is not contained in the NLL approximation. This is reflected in a
relatively large `scale uncertainty'. 
Thus one should not rely on the resummed
$R_4$ alone for a determination of $\alpha_s^\MS$.

Indeed at finite values of $\ycut$ one should match the resummed results
with the fixed-order results.  For observables that exponentiate,
a number of matching schemes have been 
defined~\cite{MatchingTh,MatchingExp} ---
$R$-matching, $\ln R$-matching, modified $R$-matching and modified 
$\ln R$-matching. 
For $R_4$, the following matching scheme 
corresponds to $R$-matching:
\be
\label{eq:MatchedRates}
R_4^{R-\rm{match}} = R_4^{\rm NLL} + 
\left[ \left(\frac{\alpha_s}{2 \pi} \right)^2 
   \left(B_4 - B_4^{\rm NLL} \right) 
+ \left(\frac{\alpha_s}{2 \pi} \right)^3 \left(C_4 - C_4^{\rm NLL} \right)
  \right] \left( 1 + \frac{\alpha_s}{\pi} \right)^{-1} ,
\ee
where the `overlap' terms $B_4^{\rm NLL}$ and $C_4^{\rm NLL}$ 
are defined by expanding $R_4^{\rm NLL}$ out in powers of $\alpha_s$, 
in analogy to \eqn{R4expansion}. 
A modified $R$-matching scheme could be defined by replacing
$L = \ln(1/\ycut)$ by $\ln(\ycut^{-1}-y_{\rm max}^{-1}+1)$ 
in $R_4^{\rm NLL}$, where $y_{\rm max}$ is the maximum kinematic 
value of $\ycut$.
This scheme would switch the resummed prediction over to the fixed-order
prediction more quickly as $\ycut$ increases, and might therefore be
more reliable at large $\ycut$, but we have not yet implemented it.
One could try to define an analog of $\ln R$-matching by
\be
\label{eq:lnRMatchedRates}
R_4^{\ln R-\rm{match}} = R_4^{\rm NLL} {B_4 \over B_4^{\rm NLL}}
 \exp\left[ {\alpha_s\over2\pi} 
   \left( - 2 + {C_4\over B_4} -  {C_4^{\rm NLL}\over B_4^{\rm NLL}} \right)
    \right] \,,
\ee
but $B_4^{\rm NLL}$ vanishes for $\ycut \sim 0.01$, so this approach fails.

 
\figmac{5}{Du}{du}{} 
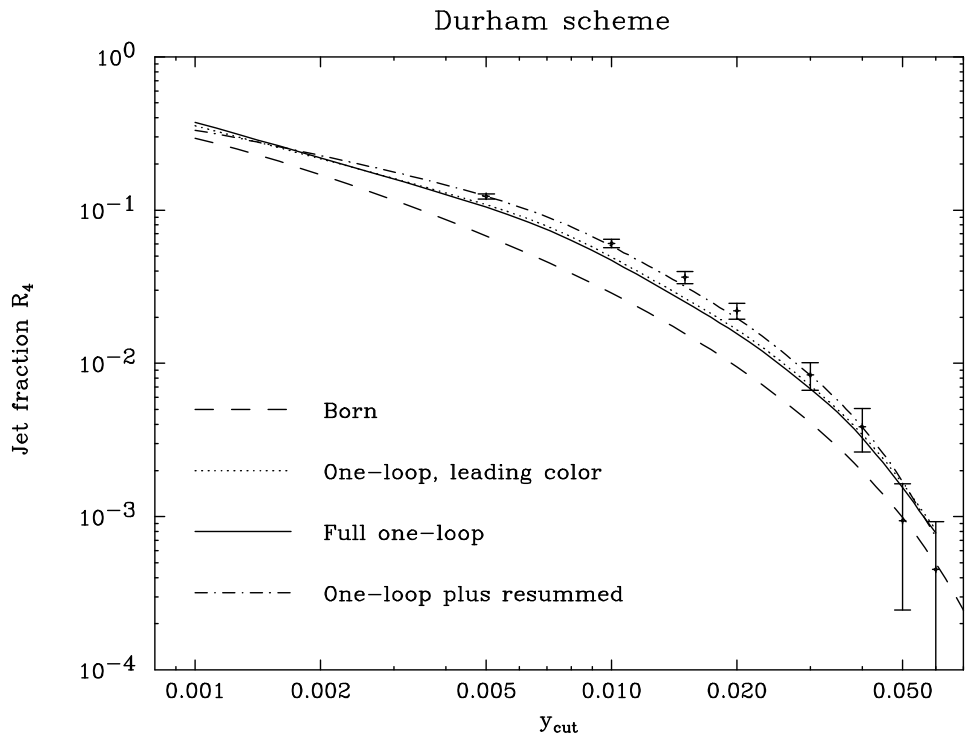
{{\small The four-jet fraction for the Durham algorithm at $\sqrt{s}=M_Z$,
illustrating the improvements to the Born term from adding successively 
the leading-color loop corrections, the subleading-color corrections, 
and the resummed corrections after matching.  
The data are from ref.~\cite{SLDdata}.
\hfill}}


We evaluate the resummed $R_4^{\rm NLL}$ using the two-loop 
formula~(\ref{eq:twoloopalpha}) for the running coupling appearing in
\eqn{NLLemit}.  To evaluate the renormalization-scale dependence of
$R_4^{\rm NLL}$ we make the substitution 
$\alpha_s \to \alpha_s + \beta_0 \ln(\mu^2/s) \alpha_s^2/2\pi$. 
In \fig{du} we show the resummed and matched prediction 
$R_4^{R-{\rm match}}$ for the Durham algorithm, together with the 
tree-level and one-loop fixed-order predictions. 
In order to illustrate once more that the subleading-color
terms are small we also show the leading-color one-loop result in
\fig{du}.

The agreement between theory and data is spectacularly good for the
resummed and matched prediction.  On the other hand, the 
`scale uncertainty' in the prediction is still sizable.
This is illustrated in \fig{dumu} where the full one-loop and the 
resummed and matched results are shown as bands. 
These bands have been obtained by varying the
renormalization scale from $ \hf M_Z < \mu < 2 M_Z $ and 
$ \third M_Z < \mu < 3 M_Z $  respectively. 
(The large scale-dependence at large $\ycut$ in the resummed and matched
prediction might be improved by a modified matching scheme.)

 
\figmac{5}{DuMu}{dumu}{} 
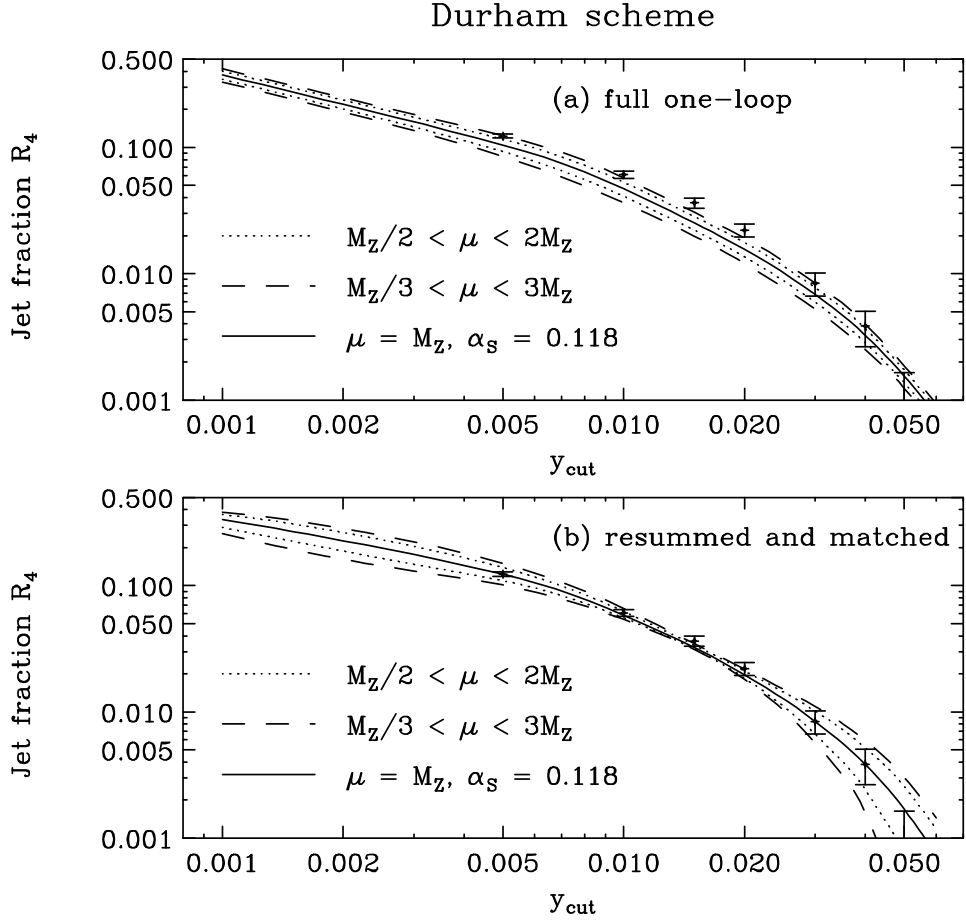
{{\small  Dependence on the renormalization scale of (a) the full one-loop
prediction and (b) the resummed and matched result, for the Durham
algorithm at $\sqrt{s}=M_Z$.
\hfill}}


\vfill\eject


\section{Conclusions}

In this article we presented the complete $\cO(\alpha_s^3)$ results
for four-jet production in electron-positron annihilation.
Generally, the NLO corrections are large and improve the agreement between 
theory and experiment considerably.  The $1/\Nc^2$-suppressed 
correction terms are indeed smaller than the leading-color terms 
by the naive factor of ten or so. 
For the Durham algorithm, after the large logarithms of $1/\ycut$ have
been resummed and the result is matched to the fixed-order prediction, 
and evaluated at the renormalization scale $\mu=\sqrt{s}$, theory
agrees remarkably well with $Z^0$ data.
Because the NLO corrections to the overall rate are so large, 
significant renormalization-scale dependence remains for both the 
fixed-order and resummed predictions, suggesting that
there are still $\sim10-20\%$ uncertainties from uncalculated 
higher-order corrections.
More precise NLO predictions are possible for normalized four-jet
distributions, for example the angles defined in
ref.~\cite{FourJetAngles},
and will be reported elsewhere~\cite{NLOAngles}.


\begin{flushleft}
{\bf\large Acknowledgement} \\
\end{flushleft}

We thank Zvi Bern, Phil Burrows and David Kosower 
for valuable conversations and suggestions.

\vskip0.2in
\par\noindent
{\it Note added in proof.}
After we submitted this manuscript, Campbell, Glover and Miller reported
on an independent calculation of the type (I) contributions to the one-loop 
virtual matrix elements for $e^+e^- \to q\qbar gg$~\cite{CGMqggq}.  
We have subsequently compared the virtual matrix elements 
used in this paper~\cite{Zqqqq,Zqggq}
to the results of refs.~\cite{CGMqggq,GloverMiller},
and we find agreement for both the four-quark and the two-quark-two-gluon
final states.   We thank J.M. Campbell and E.W.N. Glover
for providing us with numerical results from 
refs.~\cite{CGMqggq,GloverMiller}.   
Also, Nagy and Tr\'ocs\'anyi~\cite{NagyTrocsanyi}
have recently repeated our NLO calculation of $R_4$ for the
Durham, Geneva and E0 algorithms, using the matrix elements of 
refs.~\cite{CGMqggq,GloverMiller}.  They obtain general agreement with
our results, within statistical uncertainties.
However, a relatively large difference (compared to the statistical 
errors) occurring at large $\ycut$ for the Geneva 
algorithm needs further investigation.


\def\np#1#2#3  {{ Nucl. Phys. }{#1}:{#2} (19#3)}
\def\nc#1#2#3  {{ Nuovo. Cim. }{#1}:{#2} (19#3)}
\def\zp#1#2#3  {{ Z. Phys. }{#1}:{#2} (19#3)}
\def\pl#1#2#3  {{ Phys. Lett. }{#1}:{#2} (19#3)}
\def\pr#1#2#3  {{ Phys. Rev. }{#1}:{#2} (19#3)}
\def\prl#1#2#3  {{ Phys. Rev. Lett. }{#1}:{#2} (19#3)}
\def\prep#1#2#3  {{ Phys. Rep. }{#1}:{#2} (19#3)}

\end{document}